\documentclass[11pt]{article}
\usepackage{amssymb, hyperref,graphicx}
\usepackage[intlimits]{amsmath}
\title{Effect of high frequency modes of medium on an open quantum system}

\author{N. Dutta \footnote{nirupam@vecc.gov.in} and A. K. Chaudhuri\footnote{akc@vecc.gov.in}\\
\textit{Variable Energy Cyclotron Centre, 1/AF Bidhan Nagar, Kolkata 700064, India}\\
and\\
P. K. Panigrahi\footnote{pprasanta@iiserkol.ac.in}\\
\textit{Indian Institute of Science Education and Research Kolkata, Mohanpur 741252, India}
}

\begin{document}
\maketitle


\begin{abstract}
We present a method to calculate the real time effective propagator of a generic open quantum system, immersed in a medium using a wave function based framework. The medium is characterised by a set of harmonic oscillators having a continuous span of frequencies. This technique has been applied to the Caldeira-Leggett model showing that high frequency modes of the medium do not contribute towards decay of the population of states of the open system. In fact, they cause a Rabi type oscillation. Moreover, our wave function based approach provides an excellent alternative to conventional formalisms involving the density matrix.
\end{abstract}


\section{Introduction}
In physical situations where the coupling of a test system with its environment needs to be taken into account, it is customary to consider the former as an integral part of the medium in order to apply closed system dynamics to the composite unit of system plus its environment. However, this does not say much about how the medium affects the dynamics of the test system. To determine that, one may trace out the degrees of freedom of the medium and use the concept of \textit{open quantum system}. This methodology has found wide applicability in varied fields of current interest like quantum optics\cite{cohenbook1998} , quantum dissipative systems \cite{caldeira1982a,grabert1988}, adiabatic transition in chemical systems \cite{mukamel1990, tanimura1993}  and more recently, it has also been adopted to probe the effect of quark gluon plasma on heavy quark bound states \cite{borghini2011,akamatsu2011,young2010}.

Over the years, a description at the level of density matrices has constituted the key to tackling such problems. Here, evolution of the reduced density matrix of the system of interest is studied either using the Heisenberg equation in a canonical framework \cite{cohenbook1998,caldeira1982a,feynmanhibbs1965} or applying the theory of Feynman-Vernon influence functional in the context of a path integral formalism \cite{caldeira1982a,weiss1999} at zero or finite temperatures. In non pure cases  occurring usually in large systems, a knowledge of the partition function at thermal equilibrium offers much convenience in describing the medium \cite{feynmanhibbs1965}. These techniques \cite{caldeira1982a,feynman1963a,caldeira1982b}, though mostly successful in addressing the external degrees of freedom of the system, as in \cite{agarwal1971, sinha2009}, cannot, in practice, deal with the internal degrees of freedom, even in simple cases. Moreover, often a solution to the Heisenberg equation can be arrived at, only after imposing a number of stringent approximations. Added to that is the worrisome aspect of heavy computational demand of density matrix based calculation in solving for the Lindblad form \cite{lindblad1975}.

Some effective techniques \cite{reiter2012} as well as wave function based approaches \cite{dalibard1992,breuer1999,nirupamthesis,grabert1988} attempt to fill up this void in understanding open quantum systems through the use of Schr\"{o}dinger like equation. The second class of theories are computationally advantageous, requiring $N$ number of information unlike $N^2$ number of information demanded by calculations involving density matrices ($N$ is the dimension of the vector space of the system under consideration). However, in order to describe the evolution of the wave function, one needs to know the initial and final states of  the medium which requirement poses a serious difficulty in non pure cases.

In the present article, we have bypassed this dilemma by imposing certain boundary conditions that can avoid a detailed description of the medium modes. The effective propagator, obtained after tracing over the degrees of freedom of the medium, bears within itself the effect of the medium and governs the evolution of the system of interest. We observe the latter at a time scale much higher than the inherent time scale of the medium which in turn allows us to treat the dynamics of the test system as a Markovian process \cite{cohenbook1998,nirupamthesis} with no memory effect.
\section{The model}
The model system considered here is akin to the one proposed by Caldeira and Leggett (C-L) except that we have taken a generic test system instead of a harmonic oscillator. At the end of this article, this methodology has been applied to a harmonic oscillator  test system to make contact with the C-L model. The medium is composed of large number of harmonic oscillators (ideally infinite in number) based on the harmonic approximation \cite{caldeira1982a,caldeira1982b}. The oscillators in the bath do not interact among themselves but interact simultaneously with the test system. The interaction is considered to be bilinear \cite{caldeira1982a,caldeira1993} based on the assumption that it is not too strong. The Lagrangian for the the whole system (test system $+$ medium) can be written as
\begin{align}
L=L_{0}+L_{B}+L_{I} \label{eq:totlag}
\end{align}
where $L_{0}=\frac{1}{2}m_{0}\dot{x^{2}}-V$, $L_{B}=\frac{1}{2}\sum_{j}m\big(\dot{X_{j}^{2}}-\omega_{j}^{2}X_{j}^{2} \big)$ , $L_{I}=\sum_{j}\lambda(\omega_{j}) x(t) X_{j}(t) $ are the Lagrangians for the test system, bath and the interaction part
respectively. The action of the whole system is given by, 
\begin{align}
S&=S_{0} +\sum_{j}\Big[ \int_{t_{i}}^{t_{f}} dt \lambda(\omega_{j}) x(t) X_{j}(t) \nonumber \\
& +  \int_{t_{i}}^{t_{f}} dt \frac{1}{2} m\Big(\dot{X_{j}^{2}}-\omega_{j}^{2}X_{j}^{2}\Big)\Big] ,
\end{align}
$S_{0}$ is the action for the test system.


\section{Effective Propagator} 

The solution of the time dependent Schr\"odinger equation, $(H - i\frac{\partial}{\partial t}) \psi(x,t) = 0$, can in general be written as,
\begin{align}
 \psi(x,t) = \int  K(x,t;x_{i},t_{i}) \psi(x_{i},t_{i}) dx_{i}\ , \quad t>t_{i} ,
\end{align}
with the propagator $K(x,t;x_{i},t_{i}) = \langle x,t|x_{i},t_{i} \rangle = \int \mathcal{D}[x]~ e^{iS}$. For the model being studied, the propagator takes the form

\begin{align}
K 
&= \int\mathcal{D}[x]
\left(e^{iS_{0}} \prod_{j}\int \mathcal{D}[X_{j}] \right.\nonumber \\
&\left.\times \exp\left[i  \int_{t_{i}}^{t_{f}} dt 
  \left(\frac{1}{2} m\dot X_j^2
- \frac{1}{2}m\omega_{j}^{2}X_{j}^{2} 
+ J(t) X_{j}(t)\right)\right]\right).
\end{align}

Now, the path integral for the oscillators representing the medium is just a product of path
integrals of forced harmonic oscillators with an external driving force
$J(t)=\lambda x(t)$. The propagator for the $j^{th}$ oscillator in the medium is given by,

\begin{align}
\int \mathcal{D}[X_{j}] 
\exp &\left[
i \int_{t_{i}}^{t_{f}} dt  
  \left( \frac{1}{2} m \dot X_j^2
- \frac{1}{2} m \omega_j^2 X_j^2 
+ J(t) X_j(t)\right)\right] \nonumber \\
&= \langle X_{j}^{f},t_{f}|X_{j}^{i},t_{i} \rangle_{J(t)=0} 
\exp\left[\frac{-i}{2}\int_{t_i}^{t_{f}}\int_{t_{i}}^{t_{f}}dt dt^{\prime} J(t) 
D(t-t^{\prime}) J(t^{\prime})\right] .
\end{align}

When we consider interaction of all the oscillators in the medium with the test
system, the propagator of the whole system looks,
\begin{align}
K
&= \int \mathcal{D}[x]e^{iS_{0}}\prod_{j}
\langle X_{j}^{f},t_{f}|X_{j}^{i},t_{i} \rangle_{J(t)=0} \nonumber\\ 
&\qquad 
\exp\left[\frac{-i}{2}\int_{t_i}^{t_{f}}\int_{t_{i}}^{t_{f}}dt dt^{\prime} J(t) 
D_{j}(t-t^{\prime}) J(t^{\prime})\right]
\label{eq:alloscprop}
\end{align}
where $D_{j}$ satisfies the equation
\begin{equation}
m\left[\frac{d^{2}}{dt^{2}} +{\omega_{j}}^{2}\right]
D_{j}(t-t^{\prime}) =
\delta(t-t^{\prime})
\label{eq:dj}
\end{equation}
By plugging the solution of eq.\eqref{eq:dj} into eq.\eqref{eq:alloscprop} we get
\begin{align}
K
&= \prod_j \langle X_j^f,t_f|X_j^i,t_i \rangle_{J(t)=0} 
 \exp\left[\frac{\lambda^2 (x_f - x_i)^2}{4m \omega_j^3}+ O\left(\frac{1}{\omega_j^4 }\right)\right] \nonumber\\ 
& \int \mathcal{D}[x]
\exp\left[i\int_{t_i}^{t_f}dt 
\left(L_{0} +
\frac{\lambda^2 x^2 (t)}{2m \omega_j^2} \right)\right]
 \label{eq:compo}
\end{align}
Taking $\omega_{j}$ to be sufficiently large, we can neglect terms of $O\left(\frac{1}{\omega_{j}^{4} }\right)$ and higher. The above expression will then be an exact result for the composite propagator where only high frequency modes of the medium interact with the test system. We are thus finally able to separate the medium and test oscillator parts of the propagator. This is possible when the dynamics of the medium modes can be dealt with  periodic boundary conditions \cite{ramond}. This could be achieved by observing the test system above a time scale which is inverse of the lower cut off of the high frequency modes of the medium. However, the above expression falls short of being a completely general one as these boundary conditions are not applicable to all oscillatory modes of the medium. A completely general treatment of the problem with a coarse grained interaction, taking into account the contribution of all the medium modes to the dynamics of the test system can be found in \cite{nirupamthesis}. The final form of the effective propagator can be read off from eq.\eqref{eq:compo}:
\begin{align}
K_{\text{eff}}
&=
N \exp\left[\sum_j 
\frac{\lambda^2 (x_f - x_i)^2}{4m \omega_j^3}\right] \nonumber\\ 
&\quad 
\int \mathcal{D}[x]
\exp\left[i\int_{t_i}^{t_f}dt
\left(L_0+ \sum_j \frac{\lambda^2 x^2(t)}{2m \omega_j^2}\right)\right]
\label{eq:highfreqprop}
\end{align}
It is interesting to note that the effect of interactions with high frequency modes of the medium is borne entirely by the exponential outside the path integral.  Modification due to the interaction lies in the way the higher modes are distributed.
It also alters the potential of the test system by a term quadratic in $x$. This can, in general, change the form of the potential. Hence, a renormalising term must be added to the potential in eq.\eqref{eq:totlag} to counter this change \cite{caldeira1982a}:
\begin{align}
V\rightarrow V^{\prime}= V + \sum_{j}\frac{\lambda^{2}x^{2}(t)}{2m\omega_{j}^{2}}\ .
\end{align}
It is fairly straightforward to generalise the effective propagator $K_{\text{eff}}$ in eq.\eqref{eq:highfreqprop} to any test system by exploiting the fact that the path integral in eq.\eqref{eq:highfreqprop} is just the propagator of the test system. Thus, we can write
\begin{align}
K_{\text{eff}} =
N \exp \left[\frac{1}{4}\alpha(x_f - x_i)^2\right] K_0 (x_i,t_i;x_f,t_f) ,
\end{align}
where $K_{0}$ denotes the propagator of the open system and $\alpha = \sum_{j}\frac{\lambda_{j}^{2}}{m\omega_{j}^{3}}$ encodes the effect of the high frequency modes of the medium. Here, $\lambda_{j}$ is the coupling of $j^{th}$ mode of the medium with the test system. For simplicity, it can be taken to be independent of the frequency $\omega_j$ of the mode. The quantity $\alpha$ can be calculated from the distribution of the modes in the medium. This has been carried out in the following section for Ohmic and non Ohmic bath.


\section{Calculation of $\alpha$}
Taking a continuous distribution of medium modes, we shall calculate $\alpha$ using the spectral function of the medium.
\begin{align}
\rho(\omega)= \frac{2}{\pi}\sum_{j}\frac{\lambda_{j}^2}{m_{j}\omega_{j}}\delta(\omega - \omega_{j})
\end{align}
For Ohmic case $\rho(\omega)=\eta \ \omega$,
where $\eta$ determines the dissipation in the classical equation of motion of the test system (in a medium) under a potential $V$, acted upon by some external force $F_{ext}$, e.g.,
\begin{align}
M \frac{d^2}{dt^2}x + \eta \frac{dx}{dt}+\frac{dV}{dx}=F_{ext}.
\end{align}
Therefore,
\begin{align}
\alpha = \int_{\Omega_{c}}^{\Omega}\frac{\eta}{\omega}d\omega = \eta \ln(\frac{\Omega}{\Omega_{c}}) .
\end{align}
Here, $\Omega$ is the upper bound of the frequency cut off that determines the time scale of the problem. $\Omega_{c}$, on the other hand, is the lower bound of the high frequency modes, which is much larger than the characteristic frequency of the test oscillator. The value of $\alpha$ therefore depends on these two bounds and as well as on the value of $\eta$. Similarly, computation is possible for non Ohmic bath when the spectral function $\rho(\omega)\propto \omega^n$. $n$ is greater than one for super Ohmic cases and less than one for sub Ohmic cases. Analogously the quantity,
\begin{align}
\alpha = \int_{\Omega_{c}}^{\Omega} \eta \ \omega^{n-2} d\omega .
\end{align}
In the next section, we shall take a harmonic oscillator as the test system and investigate the dynamics of ground state population of the oscillator in the medium.

\section{Dynamics of population}

It is easy to ascertain from eq.\eqref{eq:highfreqprop} that the modes of the medium with a high frequency compared  to the energy of the test system do not provide any kind of damping to the classical motion of the system.  Now, let us focus on the special case where the test system is a harmonic oscillator with frequency $\Omega_T$. We prepare it initially in its ground state $|\psi_{0}\rangle$. The final state $|\psi_{f}\rangle$ can be calculated using eq. ($3$).

\begin{align}
\psi(x_{f}) = \int K_{eff}(x_{f},t_{f};x_{i},t_{i})\psi(x_{i},t_{i}) dx_{i}
\end{align}

The survival probability $\rho$ of an initial bound state evolves with time as $\rho(t)= |\langle\psi_{0}|\psi_{f}\rangle|^{2}$. The effective propagator and, hence, the ground state population dynamics depends on the lower cutoff ($\Omega_c$) of the high frequency modes of the medium that are coupled to the test system. If this lower bound is set far above the frequency of the test oscillator, then its ground state population shows an oscillatory pattern.

But as we increase the lower cutoff, the oscillation dies out and the test system remains in its initial energy eigenstate without being noticeably affected by these medium modes.

\begin{figure}[h]
\includegraphics[scale=0.3]{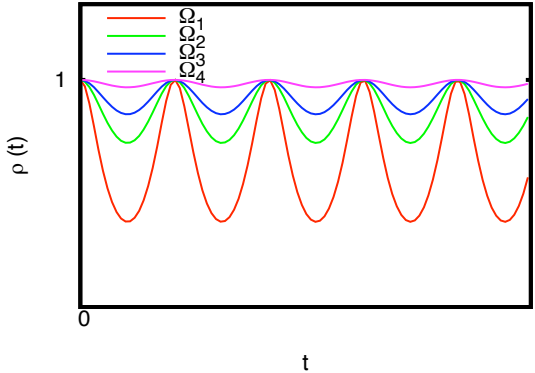}
\caption{Dynamics of survival probability of ground state population for a harmonic oscillator in an Ohmic bath for different $\Omega_c$}
\label{fig:cutoff}
\end{figure}

\begin{figure}[h!]
\includegraphics[scale=0.3]{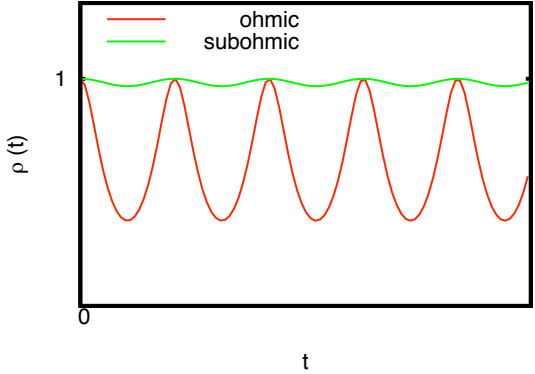}
\caption{Dynamics of ground state population of test oscillator for Ohmic and sub Ohmic bath}
\label{fig:diffbath}
\end{figure}

Fig. \ref{fig:cutoff} demonstrates this phenomenon for different values $\Omega_{1}< \Omega_{2}< \Omega_{3} <\Omega_{4}$ of the lower cutoff  $\Omega_c$. For fixed $\Omega_{c}$, $\Omega$ and $\eta$, the next plot, fig. \ref{fig:diffbath} depicts how this effect varies with the chosen medium (Ohmic and sub Ohmic).

\section{Discussion}
To sum things up, we can say that the high frequency medium modes do not contribute towards the decay of population of an open quantum system immersed in a bath. Rather, a Rabi type oscillation in the dynamics of the population is set up by them. In particular, for the harmonic oscillator test system in Caldeira Leggett model, we could illustrate that the physical aspects of this interaction is independent of the nature of the bath, though some specific details vary (fig. \ref{fig:diffbath}).

Our studies that constitute this article have been carried out using the real time effective propagator (deduced in eq.\eqref{eq:highfreqprop}) of the test system. This wave function based technique is a formal improvement over the extant theories of open quantum systems though we have not yet dealt with the entire range of frequencies of the medium. In fact, it not only is applicable to the analysis of systems having any arbitrary initial population but can also be utilised to see the transition between different states. At present, we are working on a generalisation of this cost effective methodology \cite{nirupamprep} that can take into account the influence of all possible medium modes. References \cite{dalibard1992, breuer1999} provide an alternate treatment that can include the effect of all modes without distinguishing between low and high frequencies.

Although, the phenomenon of decay of population in open systems through a non unitary evolution is well known, we expect to address various aspects of such physical systems judiciously with the help of our alternate framework, of which this letter forms a foundation.

\paragraph{Acknowledgement}
Nirupam Dutta acknowledges support from the Deutsche Forschungsgemeinschaft under grant GRK-881, IISER Kolkata and VECC, Kolkata, India.

\bibliographystyle{plain}
\bibliography{opensys}{}

\end{document}